\documentclass{edm_article}

\usepackage{balance}
\usepackage{multirow}

\begin{document}

\title{Evaluation of Fairness Trade-offs \\ in Predicting Student Success}

%
\numberofauthors{2}
\author{
\alignauthor
Hansol Lee\\
       \affaddr{Cornell University}\\
        \affaddr{Ithaca, NY, USA}\\
       \email{hl838@cornell.edu}
\alignauthor       
       Ren\'e F. Kizilcec\\
       \affaddr{Cornell University}\\
       \affaddr{Ithaca, NY, USA}\\
       \email{kizilcec@cornell.edu}
}

\maketitle


\begin{abstract}
Predictive models for identifying at-risk students early can help teaching staff direct resources to better support them, but there is a growing concern about the fairness of algorithmic systems in education. Predictive models may inadvertently introduce bias in who receives support and thereby exacerbate existing inequities. We examine this issue by building a predictive model of student success based on university administrative records. We find that the model exhibits gender and racial bias in two out of three fairness measures considered. We then apply post-hoc adjustments to improve model fairness to highlight trade-offs between the three fairness measures.
\end{abstract}



\section{Introduction}

The use of predictive models in higher education to identify at-risk students can offer advantages for the efficient allocation of resources to students. Teaching staff may direct support to struggling students early in the course, or advising staff may guide students on course planning based on model predictions. However, there is a growing concern with predictive models of this kind because they may inadvertently introduce bias~\cite{ 10.1145/3303772.3303838, 10.1145/3303772.3303791, DBLP:conf/edm/HuttGDD19, loukina-etal-2019-many, ranger2020}. For example, an unfair model may fail to identify a successful student more frequently because of their membership in a certain demographic group.

In this work, we build a course success prediction model using administrative academic records from a U.S. research university and evaluate its fairness using three statistical fairness measures: demographic parity \cite{10.1145/2783258.2783311}, equality of opportunity \cite{NIPS2016_6374}, and positive predictive parity \cite{Chouldechova2016FairPW}. Demographic parity requires an equal rate of positive predictions for different subgroups. Equality of opportunity requires that the model can correctly identify successful students at equal rates for different subgroups. Positive predictive parity requires that the proportion actually successful students out of those who receive positive predictions is the same for different subgroups. 
However, according to the impossibility results for these different statistical measures of fairness \cite{Chouldechova2016FairPW, kleinberg2016inherent}, it is not possible to satisfy any two of them at once. We therefore investigate how correcting for one fairness measure, equality of opportunity, of the student success prediction model may affect model accuracy and performance on the other two fairness measures considered.

\section{Methods}
We build a prediction model to identify students who will receive a median grade or above in one of six required courses for a given major at a U.S. research university, and evaluate its accuracy and fairness. We then alter the predictions in the post-processing step to improve fairness and evaluate the model performance again using the same criteria. 

\begin{table}[h!]
\centering
\caption{Overview of student features in the model.}
\label{tab:my-table1}
\begin{tabular}{l|l}
\textbf{Category}                                                    & \textbf{Features}                                                                  \\ \hline
\begin{tabular}[c]{@{}l@{}}Student \\ Demographics\end{tabular}      & \begin{tabular}[c]{@{}l@{}}Gender\\ First-generation status\\Racial-ethnic group\end{tabular}  \\ \hline
\begin{tabular}[c]{@{}l@{}}Academic \\ Information\end{tabular} &
  \begin{tabular}[c]{@{}l@{}}Standardized test scores \\ Standardized test scores reported\\ Previous cumulative GPA\\ Undergrad or Grad\\ Academic program \\ Academic level \\ Student major\\ Double major or not\end{tabular} \\ \hline
\begin{tabular}[c]{@{}l@{}}Top 20\\ Prior Courses\end{tabular}       & \begin{tabular}[c]{@{}l@{}}Enrolled in the course\\ Grades in each course\end{tabular} \\ \hline
\begin{tabular}[c]{@{}l@{}}Target Course \\ Information\end{tabular} & \begin{tabular}[c]{@{}l@{}}Course number\\ Offered in Fall or Spring\end{tabular} 
\end{tabular}
\end{table}

\begin{table*}[t]
\centering
\caption{Accuracy, recall, proportion of positive predictions, and precision values of the original (orig) and the fairness-corrected (fair) model for different ethno-racial groups (left) and genders (right). The group differences $\Delta$ for these four metrics indicate how fair the predictions are in terms of accuracy, equality of opportunity, demographic parity, and positive predictive parity. The statistical significance of differences is denoted by $^{***}$ for $p<0.001$, $^{**}$ for $p<0.05$ and $^{*}$ for $p<0.1$.}

\label{tab:my-table}
\begin{tabular}{cc|ccc|ccc}
 &  & \textbf{URM} & \textbf{\begin{tabular}[c]{@{}c@{}}Non-\\ URM\end{tabular}} & \textbf{$\Delta$} & \textbf{Male} & \textbf{Female} & \textbf{$\Delta$} \\ \hline
\multicolumn{1}{c|}{\multirow{2}{*}{\textbf{Accuracy}}} & orig & 0.695 & 0.735 & 0.040 & 0.694 & 0.755 & 0.061$^{**}$ \\
\multicolumn{1}{c|}{} & fair & 0.736 & 0.729 & -0.007 & 0.704 & 0.740 & 0.036 \\ \hline
\multicolumn{1}{c|}{\multirow{2}{*}{\textbf{\begin{tabular}[c]{@{}c@{}}Equality of opportunity\\(Recall)\end{tabular}}}} & orig & 0.649 & 0.851 & 0.202$^{***}$ & 0.755 & 0.860 & 0.105$^{***}$ \\
\multicolumn{1}{c|}{} & fair & 0.761 & 0.762 & 0.001 & 0.781 & 0.780 & -0.001 \\ \hline
\multicolumn{1}{c|}{\multirow{2}{*}{\textbf{\begin{tabular}[c]{@{}c@{}}Demographic Parity\\(Prop. Positive Predictions)\end{tabular}}}} & orig & 0.473 & 0.754 & 0.281$^{***}$ & 0.624 & 0.753 & 0.129$^{***}$ \\
\multicolumn{1}{c|}{} & fair & 0.556 & 0.636 & 0.080$^{**}$ & 0.645 & 0.655 & 0.010 \\ \hline
\multicolumn{1}{c|}{\multirow{2}{*}{\textbf{\begin{tabular}[c]{@{}c@{}}Positive Predictive Parity\\(Precision)\end{tabular}}}} & orig & 0.770 & 0.787 & 0.017 & 0.753 & 0.805 & 0.052$^{*}$ \\
\multicolumn{1}{c|}{} & fair & 0.767 & 0.835 & 0.068$^{*}$ & 0.753 & 0.840 & 0.087$^{**}$
\end{tabular}

\end{table*}
 
 
 \begin{figure*}
\centering
\begin{minipage}[b]{.45\textwidth}
\includegraphics[width=\linewidth]{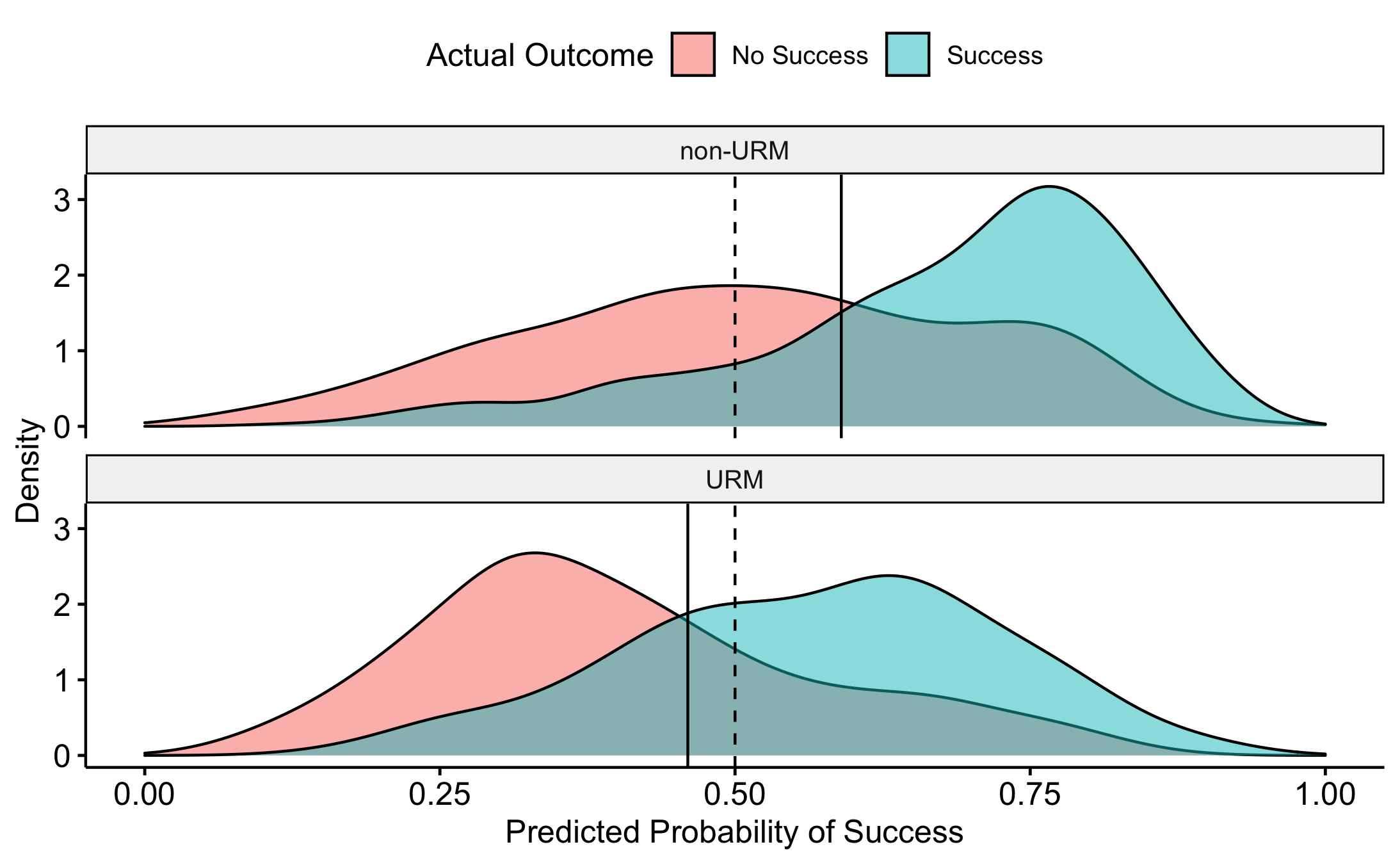}
\end{minipage}\qquad
\begin{minipage}[b]{.45\textwidth}
\includegraphics[width=\linewidth]{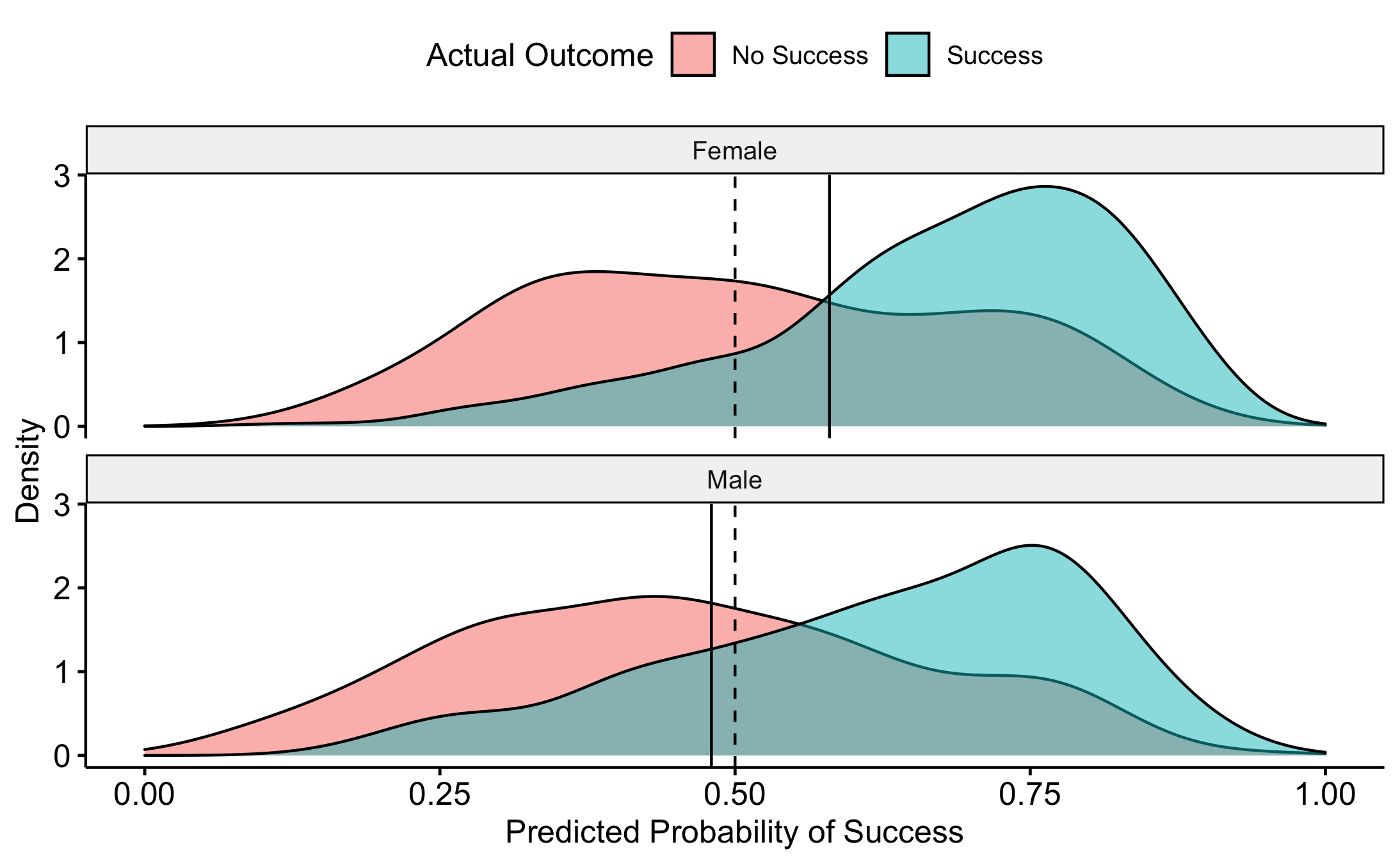}
\end{minipage}
\caption{Probability distributions of success for underrepresented minority (URM) and non-URM students (left) and for male and female students (right) estimated using a random forest model. An equal classification threshold of 0.5 (dashed line) yields higher recall performance for non-URM and female students than for URM and male students, respectively. A group-specific threshold (solid line) that is higher for non-URM and female students but lower for URM and male students satisfies equality of opportunity.}
\end{figure*}


\subsection{Data}
The data spans the Fall 2014 to Spring 2019. The student-level administrative data used to create features includes course-taking history, demographic information, and standardized test scores along with other academic information. We remove duplicates, missing course grades, courses taken multiple times by the same student, and any grades other than letter (A-F) or pass/fail grades. We impute missing standardized test scores and course grades with a placeholder value of -999 along with an indicator variable. Table~1 shows the categorization of features considered in our analysis. Some feature values with fewer than 30 instances are merged together as ``Other''. The final processed data has 5,443 rows and 56 columns. 

We focus our analysis on two binary protected attributes of students defined by their racial-ethnicity and gender. For ethnicity, we group American Indian, Black, Hawaiian or Pacific Islander, Hispanic, and Multicultural students as underrepresented minority students (URM), and Asian and White students as non-URM. For gender, we consider male students and female students.

\subsection{Model Building}
We use the most recent semester (i.e. spring of 2019) for testing, and train the model using the rest of the semesters that comprises approximately 78.4\% of the original dataset. We fit a random forest model using default settings with the randomForest function in R. We note that the training dataset is skewed toward the positive label in the training set, comprising 60.6\% of the dataset. The default settings we use include weighting each instance with the inverse of its label proportion to achieve label balance in the training set. We find that the resulting model results in an out-of-bag error of 29.36\%.

\subsection{Improving Fairness}
The original model uses the threshold value of 0.5 to determine the label based on the estimated label probabilities for each instance, as illustrated in Figure 1. To improve fairness, we pick different threshold values for each subgroup such that equality of opportunity is achieved in the testing set. The resulting group-specific threshold values are 0.48 and 0.58 for male and female groups, and 0.46 and 0.59 for URM and non-URM groups, respectively. Then we re-evaluate the resulting predictions in terms of accuracy and fairness, using a test of equal proportions to evaluate the statistical significance of group differences for each measure.

\section{Results}
We find that the overall accuracy of the resulting model on the test data is 0.73 with an f-score of 0.80. The positive label comprises 66.6\% of the test data. Table 2 shows the results of accuracy and fairness of the model using ethnicity and gender as protected attributes. We observe that the resulting model is unfair to male and URM students in terms of demographic parity and equality of opportunity, while fair in terms of positive predictive parity. 

After correcting for equality of opportunity by adjusting the classification threshold values for each group, we find that the subgroup accuracy remains similar for both groups. For URM students, the correction slightly increases accuracy from 0.695 to 0.736. In terms of fairness, we find that the correction successfully eliminated differences in equality of opportunity for both protected attributes. The correction also yields predictions that are less biased in terms of demographic parity; however, they are more biased in terms of positive predictive parity.



\section{Discussion}
Random forest models are commonly used in educational data mining and learning analytics. Here we find that an out-of-the-box random forest model violates both equality of opportunity and demographic parity for male and URM students. We note that another notion of fairness, namely positive predictive parity, is already satisfied with the original model without introducing any fairness-related interventions. This is consistent with the findings of \cite{pmlr-v97-liu19f}, which posits predictive parity as ``the implicit fairness criterion of unconstrained learning''. Based on the impossibility results, improving the fairness of the original model for any other fairness metrics (i.e. equality of opportunity or demographic parity) therefore implies that the altered model predictions will consequently have different interpretations for each student subgroup; for example, a predicted probability of student success of 60\% may be interpreted as positive for one group but negative for another.

We find that optimizing the model to satisfy equality of opportunity perpetuates unfairness in terms of demographic parity and positive predictive parity for both gender and racial-ethnic groups, consistent with the impossibility results. There is indeed a general decrease in per-group proportions of positive predictions, but this may not matter since the main goal of this student success prediction model is to correctly assign more positive predictions to successful students, not just to any students. In addition, we observe that positive predictive parity is violated. This is due to an increase in the precision values for non-URM and female students, while the values for URM and male students remained unchanged. As this does not directly lower precision for URM and male students, this can be considered a reasonable trade-off to instead achieve an alternative notion of fairness which is equality of opportunity.

We conclude that setting group-specific threshold values to achieve a certain fairness criterion itself may be considered unfair, since it means that some students will be held to a more stringent standard to achieve accurate predictions simply because of their group membership. Our findings demonstrate that different notions of fairness are in tension with each other in the context of a standard application of predictive modeling in higher education. This calls for more open discourse and careful evaluation of the potential trade-offs and desiderata around issues of fairness in the use of predictive modeling in educational applications.

\balance
\bibliographystyle{abbrv}
\bibliography{sigproc}

\end{document}